\documentclass[aps,preprint,groupedaddress]{revtex4}  

\usepackage{amssymb}
\pdfoutput=1
\pdfoutput=1
\pdfoutput=1
\pdfoutput=1
\pdfoutput=1
\usepackage{amsmath}
\usepackage{amssymb}
\usepackage{graphicx}
\usepackage{subfigure}
\usepackage{color}
\usepackage{mathrsfs}
\usepackage[dvipsnames]{xcolor}

\usepackage[breaklinks=true,colorlinks=true]{hyperref}
\hypersetup{colorlinks=true,citecolor=blue,linkcolor=blue,urlcolor=blue}

\usepackage[utf8]{inputenc}

\begin{document}
\immediate\write16{<<WARNING: LINEDRAW macros work with emTeX-dvivers
                    and other drivers supporting emTeX \special's
                    (dviscr, dvihplj, dvidot, dvips, dviwin, etc.) >>}

\title{Sine-Gordon kink lattice}

\author{E. da Hora$^{1}$}




\author{C. dos Santos$^{2}$}


\author{Fabiano C. Simas$^{3}$}


\affiliation{$^{1}$Coordenação do Curso de Bacharelado Interdisciplinar em Ciência e Tecnologia, Universidade Federal do Maranhão, 65080-805, São Luís, Maranhão, Brazil.\\
$^{2}$Centro de F\'isica e Departamento de F\'isica e Astronomia,
Faculdade de Ci\^encias da Universidade do Porto,
4169-007, Porto, Portugal.\\
$^{3}$Departamento de Física, Universidade Federal do Maranhão, Campus Universitário do Bacanga, 65085-580, São Luís, Maranhão, Brazil.
}

 
\begin{abstract}

We consider an extended model with two real scalar fields, $\phi(x,t)$ and $\chi(x,t)$. The first sector is controlled by the sine-Gordon superpotential, while the second field is submitted to the $\chi^4$ one. The fields mutually interact via a nontrivial coupling function $f(\chi)$ that also changes the kinematics of $\phi$. We briefly review the implementation of the Bogomol'nyi-Prasad-Sommerfield (BPS) prescription \cite{prasom,bogo}. We then solve the resulting BPS equations for two different interactions $f$. The first one leads to a single kink-kink configuration, while the second one gives rise to a inhomogeneous sine-Gordon kink lattice. We study the linear stability of these new solutions, focusing on their translational modes. We also explore how the strength of the mutual interaction affects the BPS profiles. In particular, we show that a homogeneous lattice with identical kinks is attained in the regime of extremely strong interactions.

\end{abstract}


\maketitle

\section{Introduction}
\label{intro}

Topological structures are usually described as solutions to highly nonlinear field models, where they emerge as a result of phase transitions \cite{masu}. These solutions have been widely investigated not only due to their intricate physical aspects, but also to their possible applications.

The simplest topological object is a kink \cite{Enz,Fin,DHN}. It appears in $(1+1)$-dimensional models with real scalar fields. Eventually, kinks can also be obtained via first-order differential equations. These equations emerge from the implementation of the BPS method \cite{prasom,bogo}. The BPS kinks saturate a well-defined lower bound the energy, being then energetically stable. Naturally, the antikink can be obtained via the $Z_2$ symmetry inherent to the original model.

In this context, the construction of a lattice as a multi-kink structure is an interesting issue that has been studied over the last years. For instance, in Ref. \cite{Gri}, the authors investigated the generation of kink-antikink pairs in a two-dimensional $\phi^4$ scenario, and an array based on these pairs was proposed in Ref. \cite{Man}. The formation of a domain-wall lattice was considered in \cite{Vacha2}, while a lattice with alternating kinks and antikinks was investigated in Ref. \cite{Vacha1}. This last configuration was also studied recently in Ref. \cite{Man2}.

Kink lattices have also been found in connection to the inhomogeneous chiral phase in the quark matter \cite{Rev}, and kink crystalline condensates, multi-kinks and twisted kink crystals in holographic superconductor \cite{Mat1,Mat2}. In particular, periodic inhomogeneous kink-like configurations are of interest in high energy and in condensed matter physics. Twisted kink crystals in the context of the Nambu-Jona-Lasinio and chiral Gross–Neveu models were also investigated \cite{Dun1,Dun2,Dun3}.

Here, a particular detail deserves attention, i.e. an analytical description is usually possible only when the lattice is formed by both kinks and antikinks. However, these profiles are known to conspire against each other. As a consequence, such a lattice is expected to be unstable.

Moreover, the complexity of the theory depends on the number of fields. In this sense, a new class of models has been recently considered \cite{blm}. In these models, the kinematics is modified to accommodate a nontrivial interaction. This modification leads to intriguing novel structures that have been used to study field profiles under geometrical constrictions \cite{balbazmar,bazmarmen,marmen}. These geometrically constrained solutions have been applied to explain experimental results in condensed matter physics \cite{poga,chdo}. Further examples can also be found in Refs. \cite{jab,hiatma,chego,thivan}.

In a very recent letter, Bazeia and Santiago have investigated the existence of novel multi-kink configurations in a model with two real scalar fields $\phi$ and $\chi$, see Ref. \cite{bs}. These fields were submitted to a superpotential given as the summation of the standard $\phi^4$ and $\chi^4$ ones. The interaction between $\phi$ and $\chi$ was assumed to be mediated by a nontrivial coupling function $f(\chi)$ that also affects the kinematics of the first. In this context, the authors have obtained an analytical array with a multi-kink profile. This particular array was recently applied to study the fermion bound states in the presence of a periodic bosonic background \cite{diosimas}.

Inspired by their work, we now go further and consider a scenario in which one of the fields is governed by the sine-Gordon superpotential. This proposal sounds interesting due to the integrable nature of the sine-Gordon theory. However, now the sine-Gordon sector $\phi$ mutually interacts with a non-integrable one submitted to the usual $\chi^4$ superpotential. As a consequence, novel effects are expected to appear. In particular, some of us have used this enlarged scenario recently to explore the sine-Gordon field under geometrical constrictions, see Ref. \cite{hora}. It is also known to support super long-range kinks with
logarithmic tails, see Ref. \cite{amm}. 

However, we now focus on a particular new result, i.e. a novel sine-Gordon solution that assumes a multi-kink profile. As we demonstrate, this new configuration is an effect of the interaction between the fields. Moreover, its main aspects get emphasized as the strength of that interaction increases. In this sense, when the interaction vanishes, the fields recover their standard (individual) profiles. On the other hand, the regime of strong interactions reinforces the non-canonical nature of the novel solutions.

In order to present our results, this manuscript is organized as follows: in Sec. \ref{secII}, we introduce the enlarged model and important conventions. The model describes a mutual interaction between the sine-Gordon field $\phi$ and a $\chi^4$ sector. The interaction is mediated by a generalizing coupling function $f(\chi)$. We briefly review the implementation of the BPS formalism. As a result, the lower bound for the energy is obtained and the BPS equations are derived. In Sec. \ref{secIII}, we particularize our investigation by studying two different types of interaction. We then obtain the corresponding BPS solutions analytically. They engender a single kink-kink profile, and a sine-Gordon kink lattice. We plot these solutions and investigate their linear stability. We also discuss how the strength of the interactions affects the field profiles. Finally, Section \ref{secIV} brings our conclusions and perspectives regarding future investigations.

\section{Generalities} \label{secII}

We consider a model with two real scalar fields, namely, $\phi(x,t)$ and $\chi(x,t)$. The potential depends on both sectors, i.e. $V=V(\phi,\chi)$. It also defines the corresponding vacuum manifold. The lagrangian density reads
\begin{equation}
\mathcal{L}=\frac{1}{2}f\left( \chi \right) \partial _{\mu }\phi
\partial ^{\mu }\phi +\frac{1}{2}\partial _{\mu }\chi \partial ^{\mu }\chi
-V\left( \chi ,\phi \right) \text{.} \label{lr1}
\end{equation}%
Here, the fields mutually interact via the coupling function $f(\chi)$. This function also affects the kinematics of $\phi$. In particular, when $f=1$, the model decouples, and the fields can be treated individually. We work in a Minkowski flat spacetime with a $(+-)$ signature. Furthermore, we consider all fields, coordinates and coupling constants dimensionless.

The equations of motion that comes from (\ref{lr1}) for $\phi(x,t)$ and $\chi(x,t)$ read
\begin{equation}
f\left( \frac{\partial ^{2}\phi }{\partial t^{2}}-\frac{\partial ^{2}\phi }{%
\partial x^{2}}\right) +\frac{df}{d\chi }\left( \frac{\partial \chi }{%
\partial t}\frac{\partial \phi }{\partial t}-\frac{\partial \chi }{\partial x%
}\frac{\partial \phi }{\partial x}\right) =-\frac{\partial V}{\partial \phi }%
 \text{,}  \label{emplr1}
\end{equation}%
\begin{equation}
\frac{\partial ^{2}\chi }{\partial t^{2}}-\frac{\partial ^{2}\chi }{\partial
x^{2}}+\frac{1}{2}\frac{df}{d\chi }\left[ \left( \frac{\partial \phi }{%
\partial x}\right) ^{2}-\left( \frac{\partial \phi }{\partial t}\right) ^{2}%
\right] =-\frac{\partial V}{\partial \chi }\text{,}
\label{emclr1}
\end{equation}
respectively. We will return to these expressions later below.

The model (\ref{lr1}) has been used to study classical fields submitted to geometrical constrictions, see Refs. \cite{blm,balbazmar,bazmarmen,marmen}. Very recently, Bazeia and Santiago have demonstrated that it also leads to a well-behaved kink crystal \cite{bs}. In that case, the superpotential was assumed to be given as a summation of the standard $\phi^4$ and $\chi^4$ ones.

In this manuscript, we go further and assume that $\phi$ is controlled by the sine-Gordon superpotential. This case is of special interest due to the integrable nature of the sine-Gordon sector. Moreover, it was successfully used by some of us in order to construct a geometrically constrained sine-Gordon profile \cite{hora}.

The theory (\ref{lr1}) supports first-order BPS solutions when the potential is given by
\begin{equation}
V\left( \chi ,\phi \right) =\frac{1}{2}\frac{W_{\phi }^{2}}{f\left( \chi
\right) }+\frac{1}{2}W_{\chi }^{2}\text{.}\label{p}
\end{equation}%
In this case, $W_{\phi}=\partial_{\phi} W$ and $W_{\chi}=\partial_{\chi} W$ are derivatives of the superpotential $W=W\left( \chi ,\phi \right)$.

In view of Eq. (\ref{p}), the static energy density related to (\ref{lr1}) can be written in the form
\begin{equation}
\varepsilon =\frac{f}{2}\left( \frac{d\phi }{dx}\mp \frac{W_{\phi }}{f}%
\right) ^{2}+\frac{1}{2}\left( \frac{d\chi }{dx}\mp W_{\chi }\right) ^{2}\pm
\frac{dW}{dx}\text{,}
\label{sumsquare}
\end{equation}%
see Ref. \cite{blm} for algebraic details.

According to the BPS prescription \cite{prasom,bogo}, the total energy that comes from (\ref{sumsquare}) is minimized when $\phi(x)$ and $\chi(x)$ satisfy
\begin{equation}
\frac{d\phi }{dx}=\pm 
\frac{W_{\phi }}{f}\text{,}\label{bpsphi}
\end{equation}%
\begin{equation}
\frac{d\chi }{dx}=\pm W_{\chi }\text{.} \label{bpschi}
\end{equation}%
These are the BPS equations inherent to the model (\ref{lr1}). Note that Eq. (\ref{bpschi}) does not depend on $f$. As a consequence, the BPS field $\chi(x)$ can be promptly obtained once $W\left( \chi ,\phi \right)$ is known. We will use this aspect later below to construct effective BPS scenarios.

The total energy of the BPS fields can then be calculated as
\begin{equation}
E_{BPS}=\int \varepsilon _{BPS}dx=\pm \Delta W\text{,}
\label{ebps}
\end{equation}%
where we have defined
\begin{equation}
\Delta W=W\left( x\rightarrow +\infty \right) -W\left( x\rightarrow -\infty
\right)\text{,}
\end{equation}
for the sake of convenience. Here,
\begin{equation}
\varepsilon _{BPS}=\pm {\frac{dW }{dx}} \text{}
\label{edbps}
\end{equation}%
represents the energy density of the BPS profiles.

Equation (\ref{bpsphi}) reveals that the BPS profile for $\phi$ depends on both $W(\chi,\phi)$ and $f(\chi)$. In other words, the solution for $\chi$ influences that for $\phi$. This is a direct consequence of the interaction mediated by $f$.

In a recent work, some of us have argued that such interaction naturally restricts the possible choices for $f$. In this sense, the coupling function is \textit{not so arbitrary}. The interested reader is referred to Ref. \cite{hora} for more details. Here, it is sufficient to highlight that all expressions that we choose below for $f(\chi)$ provide a real BPS field $\phi(x)$.

In Ref. \cite{bs}, Bazeia and Santiago have studied the formation of a kink crystal in such an extended scenario with a non-usual $f(\chi)$. The superpotential was assumed to be given as a summation between the usual $\phi^4$ and $\chi^4$ ones. For a particular interaction $f$, the authors have verified that $\phi(x)$ assumes a BPS kink-kink profile. In the sequence, they have generalized their argument, and then obtained an infinity kink lattice.

We now study the formation of such novel profiles when $\phi(x)$ is controlled by the sine-Gordon superpotential. Here, we keep the $\chi^4$ term, for the sake of simplicity. We then focus on the sine-Gordon field itself.

Note that our proposal involves an integrable field. In the absence of interaction between $\phi$ and $\chi$, the fields behave canonically. On the other hand, as the interaction gets stronger, intriguing novel aspects are expected to appear.

So, we choose the superpotential as
\begin{equation}
W\left( \phi ,\chi \right) =-\eta \cos \left[ \phi \right] +\alpha \chi
\left( 1-\frac{1}{3}\chi ^{2}\right) \text{,}
\label{w0}
\end{equation}
which leads to%
\begin{equation}
V\left( \chi ,\phi \right) =%
\frac{\eta ^{2}}{2f\left( \chi \right) }\sin ^{2}\left[ \phi \right] +\frac{1%
}{2}\alpha ^{2}\left( 1-\chi ^{2}\right) ^{2}\text{.}
\label{px0}
\end{equation}%
see Eq. (\ref{p}). The expression above indicates that $\phi$ self-interacts according to the sine-Gordon potential, while $\chi$ is controlled by the $\chi^4$ one, as desired. Here, $\eta$ and $\alpha$ are positive coupling parameters to be fixed later.

We assume that $f(\chi)$ is regular at the boundaries. As a consequence, a localized kink must satisfy
\begin{equation}
\phi _{k}\left( x\rightarrow -\infty \right) \rightarrow 0\text{ \ \ and \ \ 
}\phi _{k}\left( x\rightarrow +\infty \right) \rightarrow \pi \text{,}
\label{bcphik001a1}
\end{equation}%
\begin{equation}
\chi _{k}\left( x\rightarrow -\infty \right) \rightarrow -1\text{ \ \ and \
\ }\chi _{k}\left( x\rightarrow +\infty \right) \rightarrow +1\text{.}
\label{bcchik00a1}
\end{equation}%
These conditions are the same standard ones. In other words, the BPS fields behave canonically in the asymptotic region. The interaction mediated by $f(\chi)$ therefore does not affect the fields values attained when $x \rightarrow \pm \infty$. In this context, the effects due to the interaction are expected to occur around the center of the BPS solutions. We point out that other choices for $f$ may change the asymptotic field values. However, this discussion is beyond the scope of the present work.

The conditions (\ref{bcphik001a1}) and (\ref{bcchik00a1}) allow us to calculate%
\begin{equation}
\Delta W_{k}=2\left( \eta +\frac{2}{3}\alpha \right)\text{,}
\end{equation}
which leads to
\begin{equation}
E_{BPS,k}=+\Delta W_{k}=2\left( \eta +\frac{2}{3}\alpha \right) \text{,}\label{e00}
\end{equation}
i.e. the energy of the BPS configuration, see Eq. (\ref{ebps}). Note that this value does not depend on $f$. In this sense, the coupling between $\phi$ and $\chi$ does not affect the energy of the BPS scenario.

Still concerning the energy spectrum, the first-order Eqs. (\ref{bpsphi}) and (\ref{bpschi}) allow to rewrite the BPS energy density (\ref{edbps}) as
\begin{equation}
\varepsilon _{BPS}=\varepsilon _{\phi}+\varepsilon _{\chi}\text{,}
\end{equation}%
where we have defined
\begin{equation}
\varepsilon _{\phi}=\frac{W_{\phi }^{2}}{%
f\left( \chi \right) }=\frac{\eta ^{2}}{f}\sin ^{2}\left[ \phi \right] \text{,}\label{edp}
\end{equation}%
\begin{equation}
\varepsilon _{\chi}=W_{\chi }^{2}=\alpha ^{2}\left(
1-\chi ^{2}\right) ^{2}\text{.}\label{edc}
\end{equation}%
Here, we have also used Eq. (\ref{w0}) for $W$. Equations (\ref{edp}) and (\ref{edc}) represent the energy densities related to the BPS fields $\phi$ and $\chi$, respectively. In particular, Eq. (\ref{edp}) shows that the coupling function influences the energy distribution inherent to the sine-Gordon sector.

Now, using the superpotential (\ref{w0}), Eq. (\ref{bpschi}) provides the well-known kink solution (we have chosen the positive sign, for the sake of illustration)
\begin{equation}
\chi _{k}\left( x\right) =\tanh \left[ \alpha \left( x-x_{0}\right) \right] \text{.}
\label{schik001a1}
\end{equation}
Here, $\alpha^{-1}$ and $x_{0}$ represent the width and the position of the kink, respectively. The solution above clearly satisfies the boundary conditions (\ref{bcchik00a1}). As argued previously, this kink profile does not depend on $f$.

In the sequence, we focus on the sine-Gordon field $\phi(x)$. In view of Eq. (\ref{w0}), the BPS Eq. (\ref{bpsphi}) assumes the form%
\begin{equation}
\frac{d\phi }{dy}=\pm \eta \sin \left[ \phi \right]\text{,}
\label{ephiy0a10}
\end{equation}%
where we have introduced the new coordinate $y$. The solution reads
\begin{equation}
\phi _{k}\left( y\right) =2\arctan \left[ \exp \left( \eta \left(
y-y_{0}\right) \right) \right] \text{,}  \label{phikya1}
\end{equation}%
where $y_{0}$ is an integration constant.

It is now necessary to rewrite the above solution as a real one for $\phi_{k}(x)$. In this sense, we need to determine the relation between $y$ and the original coordinate $x$. This relation must be obtained as the solution to
\begin{equation}
\frac{dy}{dx}=f^{-1}(\chi_{k}(x))\text{,}  \label{fc0a1}
\end{equation}%
where $\chi_k(x)$ is given by the BPS profile (\ref{schik001a1}).

We notice that the solution for $y(x)$ depends on both $\chi_k(x)$ and $f(\chi_k)$. This dependence naturally imposes restrictions on $f$ itself, i.e. on the way the fields interact. The argument is that, given $\chi_k$ and $f$, Eq. (\ref{fc0a1}) must provide an analytical \textit{real} relation between $y$ and $x$. Only such a relation allows us to rewrite $\phi_k(y)$ as a \textit{manifestly real} solution for $\phi_k(x)$.

In the next Section, we focus on two different choices for $f(\chi_k)$. We explore the novel BPS profiles that they provide, and comment about their relevant aspects.

\section{BPS multi-kink configurations} \label{secIII}

We now use the first-order expressions introduced above to construct new BPS solutions. As we demonstrate, these solutions engender novel multi-kink profiles. We consider two different choices for $f(\chi_k)$. In all cases, we find a real relation $y(x)$. As a consequence, a real solution for $\phi_k(x)$ always exists.

{\subsection{Single kink-kink solution}}

We choose the interaction as
\begin{equation}
f\left( \chi \right) =\frac{1+\lambda }{1+\lambda \chi ^{-2}}\text{.}
\label{fx100a1}
\end{equation}%
Here, $\lambda$ stands for a positive coupling parameter.

The above expression generalizes the one proposed in Ref. \cite{bs}. The limit $\lambda \rightarrow 0$ leads to $f \rightarrow 1$. In this regime, the two sectors decouple, and the fields can be treated individually. On the other hand, when $\lambda \rightarrow \infty$, $f(\chi)$ behaves as $\chi^2$. In this case, the coupling between $\phi$ and $\chi$ is strong, and the analysis follows the steps proposed in \cite{bs}. The novelty is that Eq. (\ref{fx100a1}) allows the study of the effects that occur in between these two regimes, i.e. for intermediary $\lambda$.

In general, as $\chi_k$ given by (\ref{schik001a1}) interpolates between $-1$ to $+1$, even an extremely small non-vanishing $\lambda$ forces $f(\chi_k)$ to diverge at $x_0$. As a consequence, a novel BPS sine-Gordon configuration with a kink-kink profile is formed.

We now look for an explicit solution $\phi_k(x)$. In order to attain this goal, we first need to find the relation between $y$ and $x$. In view of Eqs. (\ref{schik001a1}) and (\ref{fx100a1}), Eq. (\ref{fc0a1}) becomes
\begin{equation}
\frac{dy}{dx}=\frac{1+\lambda \tanh ^{-2}\left[ \alpha \left( x-x_{0}\right) %
\right] }{1+\lambda }\text{,}  \label{dydxa1}
\end{equation}%
whose solution is
\begin{equation}
y(x)=\left( x-x_{0}\right) -\frac{\lambda }{\alpha \left( 1+\lambda \right) }%
\coth \left[ \alpha \left( x-x_{0}\right) \right] \text{.}  \label{yxa1}
\end{equation}%

\begin{figure*}[!ht]
\begin{center}
  \centering
    \includegraphics[width=1.0 \textwidth]{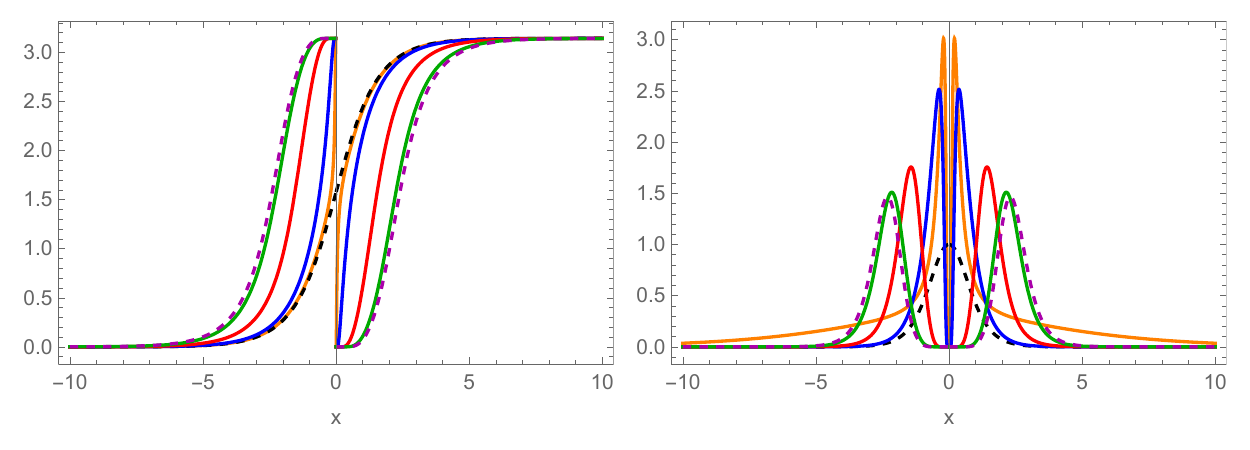}\label{fig4a}
    \vspace{-1.5cm}
  \caption{Single kink-kink solutions to $\phi_k(x)$ (left) and $\varepsilon _{\phi}(x)$ (right) for $\lambda=0$ (standard profile, dashed black line), $\lambda=0.01$ (orange line), $\lambda=0.1$ (blue), $\lambda=1$ (red), $\lambda=10$ (green) and $\lambda \rightarrow \infty$ (dashed magenta). Here, we have fixed $y_0=0$, $x_0=0$, $\eta=1$ and $\alpha=0.5$.}
  \label{fig4x}
\end{center}
\end{figure*}

Equation (\ref{phikya1}) can then be written as%
\begin{equation}
\phi _{k}\left( x\right)
=2\arctan \left[ \exp \left( \eta \left( y_{0}+\left( x-x_{0}\right) -\frac{%
\lambda }{\alpha \left( 1+\lambda \right) }\coth \left[ \alpha \left(
x-x_{0}\right) \right] \right) \right) \right] \text{.}  \label{sphik001a1}
\end{equation}%
This solution satisfies the conditions (\ref{bcphik001a1}) and therefore represents a new sine-Gordon BPS kink. Its novel profile is a clear manifestation of the nontrivial coupling mediated by $f(\chi)$. As usual, the corresponding antikink can be obtained via the $Z_2 \times Z_2$ symmetry inherent to the model (\ref{lr1}). That solution satisfies Eq. (\ref{bcphik001a1}) in the opposite spatial direction.

In Fig. \ref{fig4x} (left), we depict $\phi_k(x)$ for different $\lambda$. There, $y_0$, $x_0$, $\eta$ and $\alpha$ are fixed. For $\lambda=0$, one gets $f=1$. In this case, the mutual interaction vanishes, and the field recovers the standard sine-Gordon kink. As $\lambda$ increases, the interaction begins to play a relevant role. In this case, $f$ diverges at the center of the new solution. However, this divergence is equilibrated by the behavior of $W_{\phi}=\eta \sin{\phi}$. As a consequence, the sine-Gordon field develops a gap at $x_0$. This gap gets more and more evident with increasing $\lambda$. For sufficiently large $\lambda$ (i.e. for a strong interaction), a novel kink-kink profile appears.

It is also interesting to consider the energy distribution of the new sine-Gordon field. Given Eqs. (\ref{schik001a1}), (\ref{fx100a1}) and (\ref{sphik001a1}), the energy density $\varepsilon _{\phi}$ can be written as
\begin{eqnarray}
\varepsilon _{\phi} \left( x\right)  =&&\eta^2 \frac{1+\lambda \tanh ^{-2}\left[ \alpha \left(
x-x_{0}\right) \right] }{1+\lambda } \times   \notag \\
&&\sin^{2} \left[ 2\arctan \left[ \exp \left(
\eta \left( y_{0}+\left( x-x_{0}\right) -\frac{\lambda }{\alpha \left(
1+\lambda \right) }\coth \left[ \alpha \left( x-x_{0}\right) \right] \right)
\right) \right] \right]\text{,}
\end{eqnarray}%
see Eq. (\ref{edp}).

Figure \ref{fig4x} (right) shows $\varepsilon _{\phi}(x)$ for different $\lambda$, with $y_0$, $x_0$, $\eta$ and $\alpha$ again fixed. The solution for $\lambda=0.01$ was rescaled as $\varepsilon _{\phi} \rightarrow \varepsilon _{\phi} /4$ and $x \rightarrow x/6$, for the sake of visualization. When $\lambda=0$, $\varepsilon _{\phi}$ recovers the usual single lump profile. On the other hand, for a non-vanishing $\lambda$, the energy density assumes a two-lump configuration. These two-lump profiles are directly associated to the kink-kink solutions. It is also interesting to note how $\lambda$ controls the dimensions of these lumps. For small (large) $\lambda$, the individual kinks are more (less) localized, and the corresponding energy densities reach greater (smaller) amplitudes. These amplitudes occur at the center of each individual kink. However, these centers also move: as $\lambda$ increases, they move outwards the origin, and vice-versa.

We now study the linear stability of the novel sine-Gordon configuration. With this purpose in mind, we write the fields as usual, i.e. $\phi \left( x,t\right) =\phi _{k}\left( x\right) +\mu \left( x\right) \cos
\left( \omega t\right)$ and $\chi \left( x,t\right) =\chi _{k}\left( x\right) +\psi \left( x\right) \cos
\left( \omega t\right)$. Here, $\chi _{k}$ and $\phi _{k}$ are the BPS solutions (\ref{schik001a1}) and (\ref{sphik001a1}), respectively, while $\psi \left( x\right)$ and $\mu \left( x\right)$ are small perturbations to be determined.

Up to the first-order in $\mu$ and $\psi$, Eqs. (\ref{emplr1}) and (\ref{emclr1}) assume, respectively, the linearized form
\begin{equation}
f\frac{d^{2}\mu }{dx^{2}}+{f}_{\chi } \left(\frac{d\chi _{k}}{dx}%
\frac{d\mu }{dx}+\frac{d\phi _{k}}{dx}\frac{d\psi }{dx}\right)-%
{V}_{\phi \phi }\mu -\left( {V}_{\chi \phi }-{f}%
_{\chi \chi }\frac{d\phi _{k}}{dx}\frac{d\chi _{k}}{dx}-{f}_{\chi }%
\frac{d^{2}\phi _{k}}{dx^{2}}\right) \psi =-\omega ^{2} {f}\mu\text{,}
\label{le1a}
\end{equation}%
\begin{equation}
\frac{d^{2}\psi }{dx^{2}}-{f}_{\chi }\frac{d\phi _{k}}{dx}\frac{%
d\mu }{dx}-{V}_{\phi \chi }\mu -\left( {V}_{\chi \chi }+%
\frac{1}{2} {f}_{\chi \chi }\left( \frac{d\phi _{k}}{dx}\right)
^{2}\right) \psi =-\omega ^{2}\psi \text{.}  \label{le2a}
\end{equation}%
Here, $f$, $V$ and their derivatives must be evaluated at the BPS profiles.

In general, due to their intricate structures, Eqs. (\ref{le1a}) and (\ref{le2a}) are quite hard to solve. Even in this case, it is possible to describe the translational mode analytically. We then fix $\omega = 0$, and write the above expressions as%
\begin{eqnarray}
&&\frac{d^{2}\mu }{dx^{2}}+2\alpha \lambda \frac{\chi _{k}^{-3}\left( 1-\chi
_{k}^{2}\right) }{1+\lambda \chi _{k}^{-2}}\frac{d\mu }{dx}+\frac{2\eta
\lambda }{1+\lambda }\chi _{k}^{-3}\sin \left[ \phi _{k}\right] \frac{d\psi 
}{dx}-\frac{\eta ^{2}\left( 1+\lambda \chi _{k}^{-2}\right) ^{2}}{\left(
1+\lambda \right) ^{2}}\cos \left[ 2\phi _{k}\right] \mu  \notag \\
&=&-\frac{2\eta \lambda }{1+\lambda }\left( \frac{2\eta }{1+\lambda }\cos %
\left[ \phi _{k}\right] -\frac{\alpha \left( 3+\lambda \chi _{k}^{-2}\right) 
}{\left( 1+\lambda \chi _{k}^{-2}\right) ^{2}\chi _{k}}\left( 1-\chi
_{k}^{2}\right) \right) \chi _{k}^{-3}\left( 1+\lambda \chi _{k}^{-2}\right)
\sin \left[ \phi _{k}\right] \psi \text{,}  \label{leaaxx0}
\end{eqnarray}%
\begin{equation}
\frac{d^{2}\psi }{dx^{2}}-2\eta \lambda \frac{\chi _{k}^{-3}\sin \left[ \phi
_{k}\right] }{1+\lambda \chi _{k}^{-2}}\frac{d\mu }{dx}+\frac{\eta
^{2}\lambda }{1+\lambda }\chi _{k}^{-3}\sin \left[ 2\phi _{k}\right] \mu =2%
\left[ \frac{2\eta ^{2}\lambda ^{2}}{1+\lambda }\frac{\chi _{k}^{-6}\sin ^{2}%
\left[ \phi _{k}\right] }{1+\lambda \chi _{k}^{-2}}-\alpha ^{2}\left(
1-3\chi _{k}^{2}\right) \right] \psi \text{,}  \label{lebbxx0}
\end{equation}%
where we have also implemented (\ref{px0}) and (\ref{fx100a1}) for $V$ and $f$, respectively.

The system above is satisfied by $\psi_0 \left( x\right) =\text{sech}%
^{2}\left[ \alpha \left( x-x_{0}\right) \right]$ and%
\begin{eqnarray}
\mu_0 \left( x\right) &=&\frac{1+\lambda \tanh ^{-2}\left[ \alpha \left( x-x_{0}\right) \right] }{%
1+\lambda } \times  \notag \\
&& \sin \left[ 2\arctan \left[ \exp \left( \eta \left( y_{0}+\left(
x-x_{0}\right) -\frac{\lambda }{\alpha \left( 1+\lambda \right) }\coth \left[
\alpha \left( x-x_{0}\right) \right] \right) \right) \right] \right] \text{.}
\end{eqnarray}%

In particular, the solution for $\mu_0 \left( x\right)$ is depicted in Fig. \ref{fig0m} (left) for different $\lambda$, with the others constants fixed. The profiles vanish in the asymptotic limit, as expected. The individual solutions also present no node per period.

It is now clear that the sine-Gordon field supports well-behaved BPS solutions with a kink-kink profile. In the limit of a vanishing interaction, the enlarged model recovers the usual results. The results introduced above show how the kink-kink configuration, its energy distribution and translational mode change as the interaction becomes stronger.

In what follows, we extend our analysis and construct a lattice of sine-Gordon kinks.

{\subsection{Sine-Gordon kink lattice}}

We now look for a lattice with a multi-kink profile. In this sense, we choose the interaction%
\begin{equation}
f\left( \chi \right) =\frac{1+\lambda }{1+\lambda \left( 1+\cos \left[ \text{%
arctanh}\left( \chi \right) \right] \right) ^{-1}}\text{,}
\label{fx100a1aaax}
\end{equation}%
where $\lambda$ is again a coupling constant.

As before, this expression generalizes the results presented in Ref. \cite{bs}. The limit $\lambda \rightarrow 0$ still recovers the standard non-coupled solutions. Moreover, for $\lambda \rightarrow \infty$, $f(\chi)$ behaves as $1+\cos \left[ \text{%
arctanh}\left( \chi \right) \right]$. As we demonstrate below, this regime leads to the formation of a homogeneous kink lattice. Again, $\lambda$ controls the strength of the interaction, and it sounds interesting to consider how its values affect the lattice.

An explicit expression for $\phi_k(x)$ still depends on $y(x)$. In the present case, Eq. (\ref{fc0a1}) assumes the form%
\begin{equation}
\frac{dy}{dx}=\frac{1+\lambda \left( 1+\cos %
\left[ \alpha \left( x-x_{0}\right) \right] \right) ^{-1}}{1+\lambda }\text{,%
}  \label{dydxa1xxy}
\end{equation}%
whose solution can be verified to be%
\begin{equation}
y(x)=\frac{1}{1+\lambda }\left( x-x_{0}\right) +\frac{\lambda }{\alpha
\left( 1+\lambda \right) }\tan \left[ \frac{1}{2}\alpha \left(
x-x_{0}\right) \right] \text{.}  \label{yxa1aay}
\end{equation}%

In view of it, Eq. (\ref{phikya1}) can be rewritten as%
\begin{equation}
\phi _{k}\left( x\right) =2\arctan \left[ \exp \left( \eta \left( y_{0}+%
\frac{1}{1+\lambda }\left( x-x_{0}\right) +\frac{\lambda }{\alpha \left(
1+\lambda \right) }\tan \left[ \frac{1}{2}\alpha \left( x-x_{0}\right) %
\right] \right) \right) \right] \text{,}  \label{phikya1a0ay}
\end{equation}%
which promptly satisfies the conditions (\ref{bcphik001a1}). This solution stands for a novel sine-Gordon configuration. Here, depending on the values of $\lambda$, it may assume a multi-kink profile. This behavior is a direct consequence of the non-usual interaction defined by (\ref{fx100a1aaax}).

\begin{figure*}[!ht]
\begin{center}
  \centering
    \includegraphics[width=1.0 \textwidth]{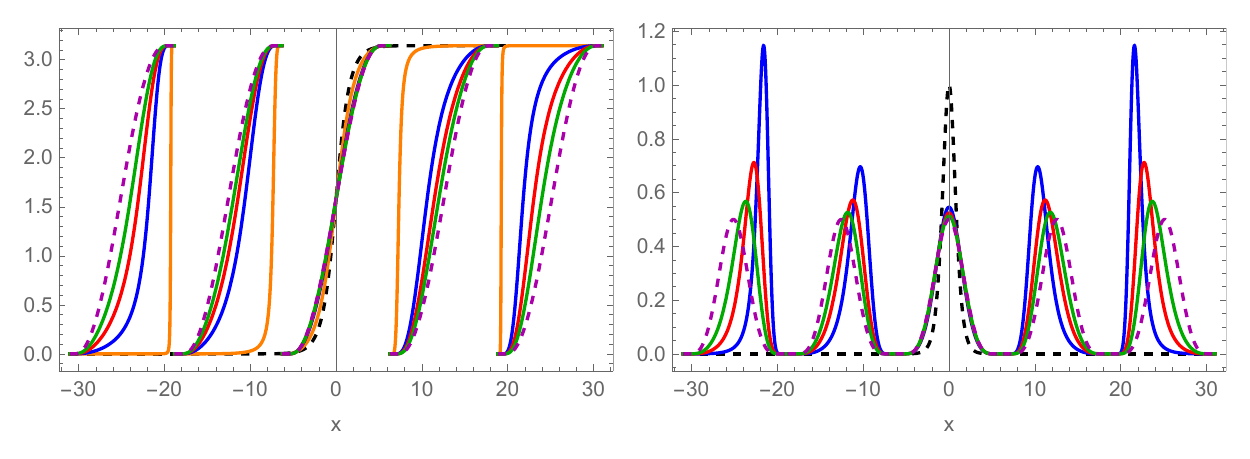}\label{fig4a}
    \vspace{-1.5cm}
  \caption{Sine-Gordon kink lattice profiles to $\phi_k(x)$ (left) and $\varepsilon _{\phi}(x)$ (right) for $\lambda=0$ (standard profile, dashed black line), $\lambda=1$ (orange), $\lambda=10$ (blue), $\lambda=20$ (red), $\lambda=40$ (green) and $\lambda \rightarrow \infty$ (dashed magenta). Here, we have fixed $y_0=0$, $x_0=0$, $\eta=1$ and $\alpha=0.5$. }
  \label{fig4xx}
\end{center}
\end{figure*}

The solution to $\phi_k(x)$ for different $\lambda$ is depicted in Fig. \ref{fig4xx} (left). Again, $y_0$, $x_0$, $\eta$ and $\alpha$ are fixed. Here, for the sake of visualization, we have chosen different values for $\lambda$ (in comparison to the previous case). As expected, $\lambda=0$ leads to the canonical [single] kink profile. In addition, $\lambda \neq 0$ gives rise to a sine-Gordon kink lattice. For small $\lambda$, the individual kinks are more localized. As this parameter increases, they become less localized, while their centers move outwards $x=0$. For sufficiently large $\lambda$, the lattice approaches the dashed magenta configuration.

As before, we consider the energy density of the kink lattice. It reads%
\begin{eqnarray}
\varepsilon _{\phi} \left( x\right)  =&&\eta^2 \frac{1+\lambda \left( 1+\cos \left[ \alpha \left(
x-x_{0}\right) \right] \right) ^{-1}}{1+\lambda } \times   \notag \\
&&\sin ^{2}\left[ 2\arctan %
\left[ \exp \left( \eta \left( y_{0}+\frac{
x-x_{0}}{1+\lambda } +\frac{\lambda }{\alpha \left( 1+\lambda \right) }\tan \left[ 
\frac{\alpha}{2} \left( x-x_{0}\right) \right] \right) \right) \right] %
\right]\text{,}
\end{eqnarray}%
whose profile is also plotted in Fig. \ref{fig4xx} (right) for different $\lambda$. The solution for $\lambda=1$ is not shown, for the sake of visualization.

In Figure \ref{fig4xx}, the single lump is the canonical result. When $\lambda \neq 0$, the energy distribution stands for a multi-lump configuration. In this case, each lump is associated with an individual kink of the lattice. As $\lambda$ varies, the dimensions of these lumps change in the same way as before, i.e. as $\lambda$ increases, the energy densities reach smaller amplitudes whose positions move outwards the origin.

An intriguing novel aspect regarding the kink lattice deserves attention. That is, for a particular $\lambda$, the individual kinks are not completely equal. So, an inhomogeneous lattice is formed. See, for instance, the blue solution in Fig. \ref{fig4xx}. Taking the kink positioned at $x=0$ as the reference, the inner ones are less localized than the outer kinks.

This asymmetry influences the energy distribution, with the inner lumps reaching smaller amplitudes. In this case, those lumps with greater amplitudes have smaller widths, and vice-versa. As a consequence, despite the asymmetry, each individual kink contributes equally for the total energy of the inhomogeneous lattice.

Figure \ref{fig4xx} shows that the asymmetry vanishes as $\lambda$ increases, and a homogeneous lattice with totally equal kinks is obtained when $\lambda \rightarrow \infty$. The conclusion is that the asymmetry is related to the strength of the interaction that originates the lattice.

Finally, we explore the stability of the lattice against small perturbations. Here, we again focus on the translational mode. Equations (\ref{le1a}) and (\ref{le2a}) can then be rewritten as%
\begin{eqnarray}
&&\frac{d^{2}\mu }{dx^{2}}-\frac{\alpha }{\lambda }\left( 1-\frac{\sin \left[
\text{arctanh}\left( \chi _{k}\right) \right] }{\Lambda _{1}}\right)
^{2}\Lambda _{1}\frac{d\mu }{dx}-\frac{\eta
^{2}}{\left( 1+\lambda \right) ^{2}}\frac{\Lambda _{1}^{2}\cos \left[ 2\phi
_{k}\right] }{\sin ^{2}\left[ \text{arctanh}\left( \chi _{k}\right) \right] }%
\mu  \notag \\
&&-\frac{\eta }{\lambda \left( 1+\lambda \right) }\frac{\left( \Lambda
_{1}-\sin \left[ \text{arctanh}\left( \chi _{k}\right) \right] \right)
^{2}\sin \left[ \phi _{k}\right] }{\left( 1-\chi _{k}^{2}\right) \sin \left[ 
\text{arctanh}\left( \chi _{k}\right) \right] }\frac{d\psi }{dx}   \notag \\
&=&\frac{\eta }{\lambda \left( 1+\lambda \right) }\frac{\left( \Lambda
_{1}-\sin \left[ \text{arctanh}\left( \chi _{k}\right) \right] \right)
^{2}\sin \left[ \phi _{k}\right] }{\left( 1-\chi _{k}^{2}\right) \sin ^{2}%
\left[ \text{arctanh}\left( \chi _{k}\right) \right] } \times  \notag \\
&&\hspace{2.0cm}\left[ \frac{2\eta }{%
1+\lambda }\Lambda _{1}\cos \left[ \phi _{k}\right] +\frac{\alpha }{\lambda }%
\Lambda _{2}\left( 1-\frac{\sin \left[ \text{arctanh}\left( \chi _{k}\right) %
\right] }{\Lambda _{1}}\right) \right] \psi \text{,}  \label{leaaxx0y}
\end{eqnarray}%
\begin{eqnarray}
&&\frac{d^{2}\psi }{dx^{2}}+\frac{\eta }{\lambda }\frac{\Lambda _{1} \sin \left[ \phi _{k}\right]}{1-\chi
_{k}^{2}}\left( 1-\frac{\sin \left[ \text{arctanh}\left( \chi _{k}\right) %
\right] }{\Lambda _{1}}\right) ^{2} \frac{d\mu }{%
dx}  \notag \\
&&-\frac{\eta ^{2}}{2\lambda \left( 1+\lambda \right) }\frac{\left( \Lambda
_{1}-\sin \left[ \text{arctanh}\left( \chi _{k}\right) \right] \right)
^{2}\sin \left[ 2\phi _{k}\right] }{\left( 1-\chi _{k}^{2}\right) \sin \left[
\text{arctanh}\left( \chi _{k}\right) \right] }\mu   \notag \\
&=&\left[ \frac{\eta ^{2}}{\lambda ^{2}\left( 1+\lambda \right) }\frac{%
\left( \Lambda _{1}-\sin \left[ \text{arctanh}\left( \chi _{k}\right) \right]
\right) ^{4}\sin ^{2}\left[ \phi _{k}\right] }{\left( 1-\chi _{k}^{2}\right)
^{2}\Lambda _{1}\sin \left[ \text{arctanh}\left( \chi _{k}\right) \right] }%
-2\alpha ^{2}\left( 1-3\chi _{k}^{2}\right) \right] \psi \text{,}
\label{lebbxx0y}
\end{eqnarray}%
where we have used (\ref{px0}) and (\ref{fx100a1aaax}). We have also defined
the auxiliary functions%
\begin{equation}
\Lambda _{1}\left( \chi _{k}\right) =\left[ 1+\lambda \left( 1+\cos \left[ 
\text{arctanh}\left( \chi _{k}\right) \right] \right) ^{-1}\right] \sin %
\left[ \text{arctanh}\left( \chi _{k}\right) \right] \text{,}
\end{equation}%
\begin{equation}
\Lambda _{2}\left( \chi _{k}\right)  =1+\left( 1+\lambda \right) \cos 
\left[ \text{arctanh}\left( \chi _{k}\right) \right]+\Lambda _{1}\left[ 2\chi _{k}\left( 1+\cos \left[ \text{arctanh}\left( \chi
_{k}\right) \right] \right) +\sin \left[ \text{arctanh}\left( \chi
_{k}\right) \right] \right] \text{,}
\end{equation}%
for the sake of convenience.

Despite its highly complicated structure, the linearized system admits the solution $\psi_0 \left( x\right) =\text{sech}%
^{2}\left[ \alpha \left( x-x_{0}\right) \right]$ and%
\begin{eqnarray}
\mu_0 \left( x\right) &=&\frac{1+\lambda \left( 1+\cos \left[ \alpha \left( x-x_{0}\right) \right]
\right) ^{-1}}{1+\lambda } \times  \notag \\
&&\sin \left\{ 2\arctan \left[ \exp \left( \eta \left( y_{0}+\frac{x-x_{0}}{%
1+\lambda } +\frac{\lambda }{\alpha \left( 1+\lambda
\right) }\tan \left[ \frac{\alpha}{2} \left( x-x_{0}\right) \right] \right)
\right) \right] \right\} \text{.} \label{mul}
\end{eqnarray}%

\begin{figure*}[!ht]
\begin{center}
  \centering
    \includegraphics[width=1.0 \textwidth]{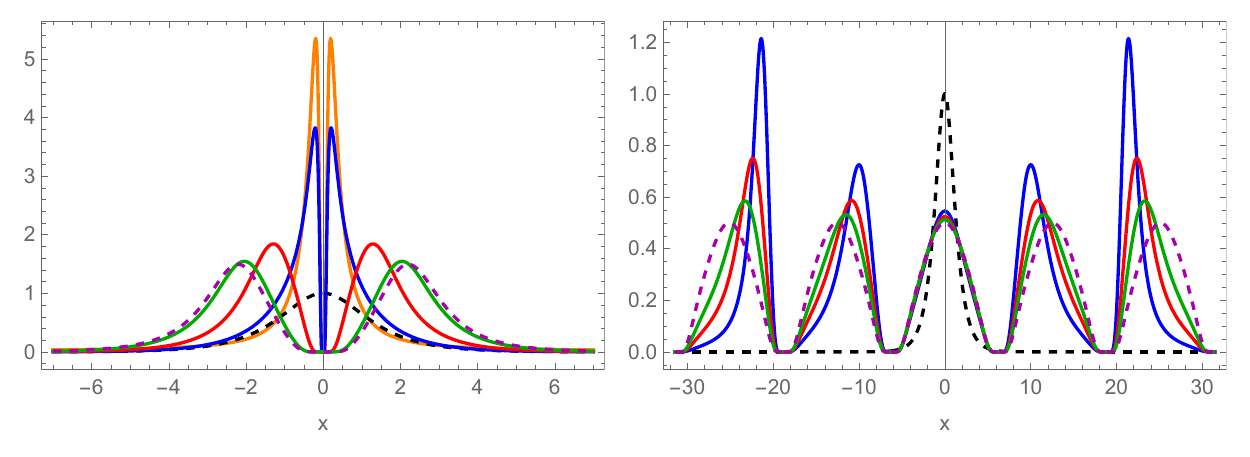}\label{fig0m}
    \vspace{-1.5cm}
  \caption{Profiles to $\mu_0(x)$ for the single kink-kink solution (left, conventions as in Fig. \ref{fig4x}) and the sine-Gordon kink lattice (right, conventions as in Fig. \ref{fig4xx}).}
  \label{fig0m}
\end{center}
\end{figure*}

Figure \ref{fig0m} (right) depicts the solution to $\mu_0 \left( x\right)$ given by Eq. (\ref{mul}). As before, we have used different values for $\lambda$. Again, the profiles present no nodes per period.

The conclusion is that a sine-Gordon kink lattice can be attained as a result of a mutual interaction. The individual kinks that form it are not equal in general, and a homogeneous lattice with identical kinks is obtained in the limit of extremely strong interactions. Despite this asymmetry, each kink contributes equally to the energy of the inhomogeneous lattice. Finally, the existence of a translational mode with no node per period can be verified analytically.

\section{Summary and perspectives} \label{secIV}

We have studied a kink lattice formed by a sine-Gordon field. With this aim in mind, we have considered an enlarged model with two real scalar sectors, i.e. $\phi(x,t)$ and $\chi(x,t)$, see Eq. (\ref{lr1}). We have assumed that these fields interact via a nontrivial coupling function $f(\chi)$ that also changes the kinematics of $\phi$. As we have demonstrated, it is precisely this interaction that supports the lattice.

We have briefly reviewed the implementation of the BPS prescription. As a result, we have obtained the first-order equations (\ref{bpsphi}) and (\ref{bpschi}) to both $\phi$ and $\chi$, respectively.

We have assumed that $\phi$ is controlled by the sine-Gordon superpotential, while $\chi$ is submitted to the $\chi^4$ one, see Eq. (\ref{px0}). Also, we have noted that the first-order Eq. (\ref{bpschi}) does not depend on $f$, from which we have written its BPS kink solution $\chi_k(x)$ promptly, see Eq. (\ref{schik001a1}). 

Furthermore, given the dependence of the sine-Gordon solution on $f$, we have divided our work into two different cases. We have proposed them as generalizations of the ones successfully studied in Ref. \cite{bs}. In this sense, we have introduced a real parameter $\lambda$ that controls the strength of interaction between $\phi$ and $\chi$.

In this context, we have first obtained an original BPS sine-Gordon solution that engenders a single kink-kink profile, see Eq. (\ref{sphik001a1}). We have verified that this novel solution recovers the canonical result as the interaction vanishes (i.e. when $\lambda=0$), see Fig. \ref{fig4x}. Moreover, we have calculated its translational mode analytically.

We have generalized the above idea, and then found an exact sine-Gordon field that now describes a kink lattice, see Eq. (\ref{phikya1a0ay}). We have observed that its individual kinks are not equal in general. However, despite this asymmetry, we have argued that they contribute equally to the energy of the inhomogeneous lattice. Also, we have depicted the corresponding profiles, and concluded that the asymmetry is related to the strength of the coupling between $\phi$ and $\chi$, see Fig. \ref{fig4xx}. We have pointed out that the asymmetry vanishes in the limit of extremely strong interactions (i.e. $\lambda \rightarrow \infty$), and a homogeneous lattice with totally equal kinks emerges. Finally, we have described analytically the translational mode of the lattice.

It is interesting to note that the coupling (\ref{fx100a1}) can be obtained from the generalized one%
\begin{equation}
f(\chi) =%
\frac{1+ \lambda}{1+\lambda \chi^{2n}} \text{,} \label{gene}
\end{equation}%
with $n \in \mathbb{Z}$. Here, $n=0$ leads to $f=1$, i.e. no coupling. In addition, when $n=+1$, the mutual interaction gives rise to the two-kink profiles explored in \cite{hora}, while $n=-1$ leads to the novel multi-kink configurations introduced in the present manuscript.

In this sense, Eq. (\ref{gene}) maps different well-established scenarios simultaneously. The same argument can be applied to the coupling (\ref{fx100a1aaax}). So, both cases can be seen as particular examples of a more complex family of interactions that therefore deserves a detailed study. Furthermore, the investigation regarding the collisions between these multi-kink structures is still an unsolved issue. These ideas are currently under consideration, and we expect relevant contributions to be presented in a future manuscript.


\section*{Acknowledgements}

E. H. thanks the Centro de Física and the Departamento de Física e Astronomia of University of Porto for their hospitality during the realization of this work. F. C. S. thanks the support from Coordenação de Aperfeiçoamento de Pessoal de Nível Superior - CAPES (Brazilian agency, via a PhD scholarship) - Finance Code 001.



\end{document}